\documentclass[foma;,3p,10pt]{elsarticle}

\usepackage{amssymb}

\usepackage{lineno}

\usepackage{graphicx}
\usepackage{amsmath}
\usepackage{subfig}
\usepackage{cases}
\usepackage{natbib}
\usepackage{multibib}
\newcites{supp}{Supplementary References}

 \def\vector#1{\mbox{\boldmath $#1$}}

\journal{arXiv}

\makeatletter
\def\@author#1{\g@addto@macro\elsauthors{\normalsize%
    \def\baselinestretch{1}%
    \upshape\authorsep#1\unskip\textsuperscript{%
      \ifx\@fnmark\@empty\else\unskip\sep\@fnmark\let\sep=,\fi
      \ifx\@corref\@empty\else\unskip\sep\@corref\let\sep=,\fi
      }%
    \def\authorsep{\unskip,\space}%
    \global\let\@fnmark\@empty
    \global\let\@corref\@empty  
    \global\let\sep\@empty}%
    \@eadauthor={#1}
}
\makeatother

\makeatletter
\def\ps@pprintTitle{%
   \let\@oddhead\@empty
   \let\@evenhead\@empty
   \let\@oddfoot\@empty
   \let\@evenfoot\@oddfoot
}

\begin{document}

\begin{frontmatter}


\title{Regressing bubble cluster dynamics as a disordered many-body system}
\author{Kazuki Maeda\corref{cor1}}
\ead{kemaeda@stanford.edu}
\cortext[cor1]{Corresponding author}




\address{Center for Turbulence Research, Stanford University\\
481 Panama Mall, Stanford, CA 94305, USA}

\author{Daniel Fuster}
\address{Sorbonne Universit\'e, CNRS\\
UMR 7190, Institut Jean Le Rond D'Alembert,  F-75005 Paris, France}

\begin{abstract}
The coherent dynamics of bubble clusters in liquid are of fundamental and industrial importance and are elusive due to the complex interactions of disordered bubble oscillations.
Here we introduce and demonstrate unsupervised learning of the coherent physics by combining theory and principal component analysis.
From data, the method extracts and quantifies coherent dynamical features based on their energy.
We analyze simulation data sets of disordered clusters under harmonic excitation. Results suggest that the coherence is lowered by polydispersity and nonlinearity but in cavitating regimes underlying correlations can be isolated in a single cohererent mode characterized by mean-field interactions, regardless of the degree of disorders.
Our study provides a valuable tool and a guidance for future studies on cavitation and nucleation in theory, simulation, and experiments.
\end{abstract}




\end{frontmatter}


\section{Introduction} 


Bubble clusters nucleate when the liquid pressure rapidly falls below a certain threshold.
These clusters coherently oscillate and violently collapse to cause extreme energy concentration that leads to various critical consequences and use in applications as diverse as hydraulic machines \citep{Plesset55,Morch80}, medical ultrasound \citep{Ikeda06,Pishchalnikov11,Movahed16}, surface cleaning \citep{Verhaagen16,Yamashita19}, chemical synthesis \citep{Suslick99,Cairos17}, and bio-inspired devices \citep{Tang19}.
Characterizing the dynamics is a challenge due to the complex interactions of bubbles involving disorders and stochasticity.
Nuclei are typically micro-sized and polydisperse and randomly distributed.
Their rapid oscillations are, except in controlled experiments \citep{Bremond06}, practically not measurable.
Molecular and hydrodynamics simulations can provide detailed insights into nucleation \citep{Angelil14,Gallo21}, while their time- and spatial scales have not reached those of practical cluster oscillations.
Analyses have been made on the interaction dynamics in various regimes \citep{Brennen14}, yet no common knowledge has been established if the many-body coherence globally exists and if so scaling is possible, beyond the consensus that polydispersity induces strong disorders.

For the past decades, the Rayleigh-Plesset (R-P) equation and its variations have been actively explored to investigate the dynamics of single bubbles \citep{Plesset49,Plesset77}. Relatively few studies addressed the theory of clusters.
By using mean field approach to interacting bubbles modeled by the R-P equation, d'Agostino and Brennen \citep{dAgostino89} derived a scaling parameter that dictates the coherent oscillations of nearly equilibrium, monodisperse clusters, the so-called "cloud interaction parameter".
Zeravcic and co-workers \citep{Zeravcic11} used the coupled R-P equations and identified disorders represented by the Anderson localization of acoustic energy in polydisperse, lattice-like clouds under weak excitation.
With the aid of computation, we have recently extended the interaction parameter to the non-equilibrium, cavitating clusters under strong excitation by combining the mean-field and the explicit approaches \citep{Maeda19,Maeda18a,Maeda21}.
To recall, we scale the kinetic energy of liquid induced by $N$ oscillating bubbles as
\begin{equation}
K \sim 2\pi\rho_l R^3_{b,c}\dot{R}^2_{b,c}(1+B_d),\label{eqn:1}
\end{equation}
where $B_d$ is the extended parameter: $B_d = NR_{b,c}/R_C$, and $R_{b,c}$ and $\dot{R}_{b,c}$ denote the characteristic (reference) bubble radius and velocity, $R_C$ denotes the radius of the cluster, and $\rho_l$ is the liquid density.
For bubbles out of equilibrium, $R_{b,c}$ is expressed as $R_{b,c}=\overline{\langle R_{i}(t) \rangle }$, where $\langle \cdot \rangle$ and $\overline{(\cdot)}$ denote time average during a period in which bubble dynamics are statistically stationary and the mean value about the bubbles in the cluster, respectively.
Despite rather audacious averaging, $B_d$ was found to control well both the coherent dynamics of clusters and their acoustic emission.
Overall, previous studies indicate that the coherence can depend on both polydispersity and nonliearity in a non-separable manner, posing perplexing questions about the universality of the scaling.

Theoretical characterization of the nonlinear dynamics of disordered many-body systems is in general not a simple task. Meanwhile, greater computing power has enabled learning physics by analysing big data.
Principal component analysis (PCA) is a powerful method for unsupervised learning which has seen recent success in characterizing the coherent physics of many-body and high-dimensional systems in fields including quantum information and fluid dynamics \citep{Lloyd14,Taira17, Milano02}.
PCA extracts dominant states and dynamical features as well as their amplitudes, such as coherent quantum state and turbulent structures, as the eigenmode and the variance of the co-variance matrix of measurement and simulation data. The variance is designed consistent with the norm induced by an energetic inner product of the system (e.g., kinetic energy) \citep{Lall99,Rowley05}.

\begin{figure*}[t!]
    \centering
     \includegraphics[width=110mm,trim=0 0 0 0, clip]{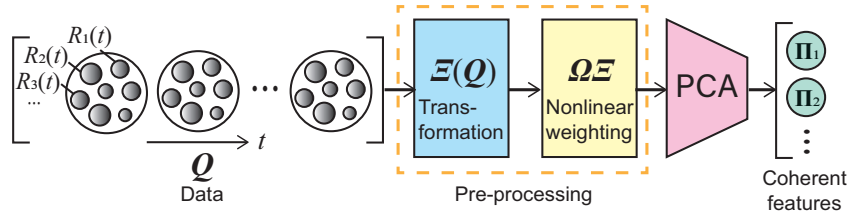}
     \caption{Schematics of the feature extraction by PCA from Lagrangian bubble dynamics data, $\vector{Q}$, after pre-processing. The variance of the resulting features is consistent with the interaction potential energy predicted by the Rayleigh-Plesset equation.}
     \label{f:scheme}
\end{figure*}
Here, we propose and demonstrate a method for the unsupervised learning of the coherent bubble cluster dynamics by combining theory and PCA.
Direct application of standard PCA to Lagrangain bubble dynamics data is not physically meaningful as the extracted features do not account for the interaction potential.
To make PCA physically meaningful, we introduce strategic pre-processing of the data prior to PCA such that the PC-variance becomes theoretically consistent with the interaction energy modeled by the R-P equation.
Analysing simulation data sets of clusters, we show that the PCA can systematically extract not only coherent but also incoherent features whose magnitudes are respectively measured by the PC-variance and the von Neumann entropy. We discover that the coherence is lost by disorders induced by polydispersity and nonlinearity, while under strong excitation the underlying correlations are globally isolated in a single coherent feature scaled by $B_d$, regardless of the disorders.

For clusters modeled by the coupled R-P equations, observable variables are bubbles' radial velocities and radii, $R$ and $\dot{R}$.
Assume data matrices contain $N_t$ observations with a constant temporal interval:
$\vector{Q}=[\vector{q}_1,\vector{q}_2,\dots,\vector{q}_{Nt}]$ and $\vector{R}=[\vector{r}_1,\vector{r}_2,\dots,\vector{r}_{Nt}]$,
where $\vector{q}_k$ and $\vector{r}_k$ denote the radial velocities and the radii of the $N$ bubbles at time $t_k$:
$\vector{q}_k=[\dot{R}_1(t_k),\dot{R}_2(t_k),\dots,\dot{R}_N(t_k)]^T$ and
$\vector{r}_k=[R_1(t_k),R_2(t_k),\dots,R_N(t_k)]^T$.
A standard PCA can be conducted through, for instance, the singular value decomposition (SVD) of $\vector{Q}$: $\vector{Q}=\vector{U}\vector{\Sigma}\vector{X}^*$ \citep{Bbdi10,Jolliffe16}.
The $i$-th PC (feature) is stored in $\vector{\Pi}_{i}=\vector{U}\vector{\Sigma}_{i}$, where $\vector{\Sigma}_{i}$ keeps the $i-th$ largest singular value and zero elsewhere.
However, corresponding PC-variances do not measure the energy.

Now we revisit eq.(\ref{eqn:1}).
$K$ can be explicitly expressed as
\begin{equation}
K = 2\pi\rho_l\sum^N_{i=1} \left[R_i^3\dot{R}_i^2+\sum^N_{j\ne i}\frac{R_i^2R_j^2 \dot{R}_i \dot{R}_j}{r_{ij}}\right]+(H.O.T),\label{eqn:2}
\end{equation}

where $R_i$, $\dot{R}_i$, and $r_{ij}$ are the radius and the radial velocity of bubble $i$, and distance between the centers of bubble $i$ and $j$, respectively.
The second term in the bracket represents the contribution of the long-range interactions.
At each instant, $K$ can be expressed as a weighted inner product of $\vector{q}$: $K=\vector{q}^T\vector{T}\vector{q}=(\vector{W}\vector{q})^T(\vector{W}\vector{q})$, where
\begin{equation}
T_{ij}(\vector{r}_k) =
\begin{cases}
\phantom{-} 2\pi\rho R^3_i(t_k) & i=j,\\
2\pi\rho \frac{R^2_i(t_k)R^2_j(t_k)}{r_{ij}} & i\ne j
\end{cases}
\end{equation}
and the weight matrix, $\vector{W}$, can be obtained through the Cholesky factorization of $\vector{T}(\vector{r})$: $\vector{T}(\vector{r})=\vector{W}\vector{W}^*$. This expression suggests the weighted data, $\vector{W}\vector{Q}$, is appropriate as the input for PCA. However, $\vector{T}(\vector{r})$ and $\vector{W}$ are time-dependent, and incompatible with PCA requiring constant weight.

We overcome this obstacle by pre-processing the data. The schematic is shown in Fig. \ref{f:scheme}. We transform $(\vector{q},\vector{r})$ into new variables $(\vector{\xi},\vector{\eta})$ and find a corresponding constant weight matrix, namely $\vector{\Omega}$, such that the PCA of weighted data, $\vector{\Omega}\vector{\Xi}$, becomes physically meaningful. $\vector{\Xi}$ is the transformed data matrix: $\vector{\Xi}=[\vector{\xi}_1,\vector{\xi}_2,...,\vector{\xi}_{N_t}]$.
Such $(\vector{\xi}, \vector{\eta}, \vector{\Omega})$ can be identified by considering the R-P equation.
First, we consider a single bubble. 
By transformation $(\xi,\eta)=(R\dot{R}, R)$, the R-P equation can be expressed as
\begin{equation}
    \dot{\xi}=-\frac{1}{2}\left(\frac{\xi}{\eta}\right)^2+\frac{p_b-p_\infty(t)}{\rho},\label{eqn:x}
    \hspace{3mm} \dot{\eta}=\frac{\xi}{\eta}.
\end{equation}
$p_b$ and $p_\infty$ are the pressure inside the bubble and that in infinity, respectively. Next, we project the system. Given data, a time average can be obtained for $\eta$ as $\langle \eta\rangle$.
Using $\langle \eta\rangle$, eqs.(\ref{eqn:x}), we project the system onto a space spanned by a set of new variables $(\hat{\xi},\hat{\eta})$.
\begin{equation}
    \dot{\hat{\xi}}=-\frac{1}{2}\left(\frac{\hat{\xi}}{\langle \eta\rangle}\right)^2+\frac{p_b-p_\infty(t)}{\rho},\label{eqn:x2}
    \hspace{3mm} \dot{\hat{\eta}}=\frac{\hat{\xi}}{\hat{\eta}}.
\end{equation}
Although $\hat{\xi}$ and $\hat{\eta}$ are partially decoupled, this system models the dynamical features of the R-P equation including bifurcation.
Physically speaking, the rapid change of $\eta$ has a relatively small influence on the evolution of $\xi$ at the timescale of (statistically) stationary bubble oscillations.
This projected system informs that $K$ can be approximated using $\langle \eta\rangle$: $K=2\pi\rho \dot{R}^2R^3=2\pi\rho \xi^2\eta\approx 2\pi\rho \hat{\xi}^2\langle \eta\rangle$.
Now we can apply the same procedures to clusters.
$K$ can be approximated for $\vector{\xi}=[\xi_1, \xi_2,...,\xi_N]^T$ and $\vector{\eta}=[\eta_1, \eta_2,...,\eta_N]^T$ as
\begin{equation}
K\approx\vector{\xi}^T\vector{P}(\langle\vector{\eta} \rangle)\vector{\xi}\label{eqn:K}
\hspace{0.5em}\mathrm{and}\hspace{0.5em}
P_{ij}(\langle\eta\rangle) =
\begin{cases}
\phantom{-} 2\pi\rho \langle \eta_i\rangle & i=j,\\
2\pi\rho \frac{\langle \eta\rangle_i \langle \eta\rangle_j}{r_{ij}} & i\ne j.
\end{cases}
\end{equation}
Now, the constant $\vector{\Omega}$ is obtained through $\vector{P}(\langle \vector{\eta}\rangle)=\vector{\Omega}\vector{\Omega}^*$.
The PC decomposition is performed as $\vector{\Omega}\vector{\Xi}=\vector{U}_{\xi}\vector{\Sigma}_{\xi}\vector{X}_{\xi}^*$.
The $i$-th PC is stored in $\vector{\Pi}_{\xi,i}=\vector{U}_{\xi}\vector{\Sigma}_{\xi,i}$. We denote the $i$-th largest singular value of $\vector{\Sigma}_{\xi}$ as $\sigma_i$.
The degree of coherence can be measured by $\hat{\sigma}_1^2=\sigma_1/\mathrm{tr}(\vector{\Sigma}^2)$, the ratio of the first PC-variance to the total energy.

We introduce two additional key measures to characterize the coherent physics. First, to quantify the degree of incoherence through PCA, we define the von Neumann entropy of the co-variance matrix of $\vector{\Omega}\vector{\Xi}$, $\hat{S}_{vN}$. $\hat{S}_{vN}$ is a quantum analogue of the Shannon entropy and is used to quantify the entanglement of states from a density matrix \citep{Kitaev06}. Here, ${S}_{vN}=\sum-\hat{\sigma}^2_k(\mathrm{log}\hat{\sigma}^2_k)$.
We normalize ${S}_{vN}$ by $\mathrm{ln}_2(N)$: $\hat{S}_{vN}=S_{vN}/\mathrm{ln}_2(N)$.
When bubbles are in perfect correlation, we expect to excite only the first PC capturing the entire energy of the system: $\hat{\sigma}^2_1=1$ and $\hat{S}_{vN}=0$. In contrast, when the energy is equi-partitioned into all PCs, $\hat{\sigma}^2_1=1/N$ and $\hat{S}_{vN}=1$.
Second, we define the coherence measure $C:C=(\sigma^2_1/\sigma'^2_1-1)/B_d$. $C$ quantifies the degree to which the coherent interactions represented by the first PC is controlled by $B_d$.
$\sigma'^2_1$ is a portion of the first PC-variance excluding the contribution of the interaction potential, obtained from PCA of $\vector{\Xi}$ weighted by an alternative matrix $\vector{\Omega}'$.
$\vector{\Omega}'$ is defined through $\vector{P}'(\langle \vector{s}\rangle)=\vector{\Omega}'\vector{\Omega}'^*$, where $\vector{P}'$ includes only the diagonal entries of $\vector{P}$: $P'_{ii}(\vector{s}_k) = 2\pi\rho\langle \eta_i\rangle$.
The corresponding inner product, $\vector{\xi}^T\vector{P}'\vector{\xi}$, represents the energy excluding the interactions.
In the limit of perfect correlation, $\sigma^2_1=\langle K\rangle\approx\sigma'^2_1(1+B_d)$ (see eq. (\ref{eqn:1})) and $C\approx 1$. This condition is typically realized in monodisperse clusters under weak (linear) oscillations \citep{dAgostino89}.

We neglect bubble's translation and deformation; their effects are short-range and can be less dominant than volumetric oscillations for the coherent physics. PCA can nevertheless incorporate dynamical variables controlling the bubble's position \citep{Ilinskii07} and the shape \citep{Murakami20}. Of course, the models able to capture these effects are usually computationally expensive and the generation of database may become prohibitive.

To verify and demonstrate PCA, we use data sets of spherical clusters excited by 40 cycles of harmonic pressure excitation.
Each cluster contains $O(10-10^3)$ bubbles with their initial radii following log-normal distributions with a reference radius of $R_{ref}=O(10)$ $\mu$m; $\mathrm{ln}(R_0/R_{ref})=N(0,\sigma_d)$, where $\sigma_d$ is the polydispersity measure \citep{Maeda19}.
The bubbles are randomly distributed in the spherical region with a specified cluster radius.
Similar parameters of clusters were previously simulated to compare with experiments \citep{Maeda19, Maeda21}.
The sparsity of bubbles is characterized by $B_0$, the equilibrium value of $B_d$.
The excitation frequency is 500 kHz unless noted, near the adiabatic resonant frequency of the reference bubble.
The pressure amplitude, $A$, is normalized by $1.0$ atm.
For data generation, we use mesh-free, coupled Keller-Miksis equations by modifying previous methods \citep{Keller80,Fuster11}.
Details are provided in supplementary material \citep{Cloud}. For PCA, the method for data generation can incorporate experimental measurements as well as other numerical approaches. The cluster's shape can be arbitrary.

\begin{figure}[t]
    \centering
    \includegraphics[width=68mm,trim=0 0 0 0, clip]{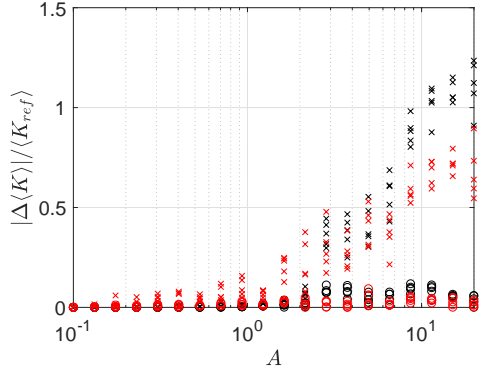}
     \caption{Relative error of the mean kinetic energy of clusters obtained by the two approximations, $(\times)$:$K\approx\vector{q}^T\vector{T}(\langle \vector{q} \rangle)\vector{q}$ and $(\circ)$:$K\approx\vector{\xi}^T\vector{P}(\langle \vector{\eta}\rangle)\vector{\xi}$, against the excitation pressure amplitude. Results for two polydispersities (black: $\sigma_d=0.1$, red: $0.7$) are shown.}
     \label{f:Kcomp}
\end{figure}
To show the effectiveness of the pre-processing, in fig. \ref{f:Kcomp}, we plot the relative errors of the time-averaged kinetic energy of sparse and dense clusters with $B_0=0.5$ and $5.0$, computed using the original and transformed variables ($\vector{q}^T\vector{T}(\langle\vector{r}\rangle)\vector{q}$ and $\vector{\xi}^T\vector{P}(\langle\vector{\eta}\rangle)\vector{\xi}$) against the excitation amplitude. At $A<10^{-1}$, the error is nearly zero for both approximations. At $A>10^{-1}$, at which bubble dynamics become nonlinear, the error grows with $A$ for the former, while it remains small for the latter. This result confirms the improved approximation by the pre-processing.

\begin{figure*}
    \centering
     \subfloat[]{\includegraphics[width=54mm,trim=5 0 35 0, clip]{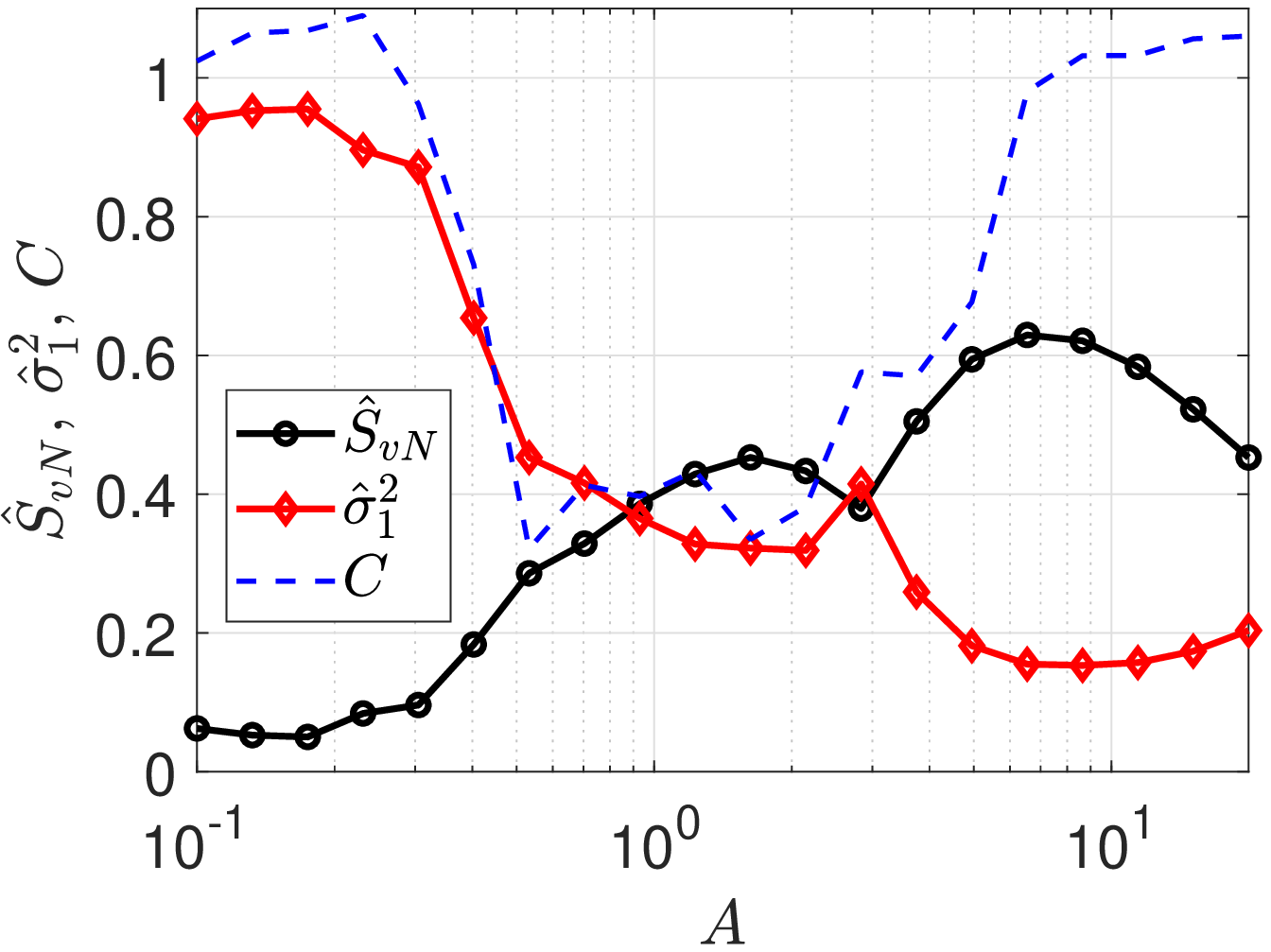}}\\
     \subfloat[]{\includegraphics[width=54mm,trim=-5 0 30 0, clip]{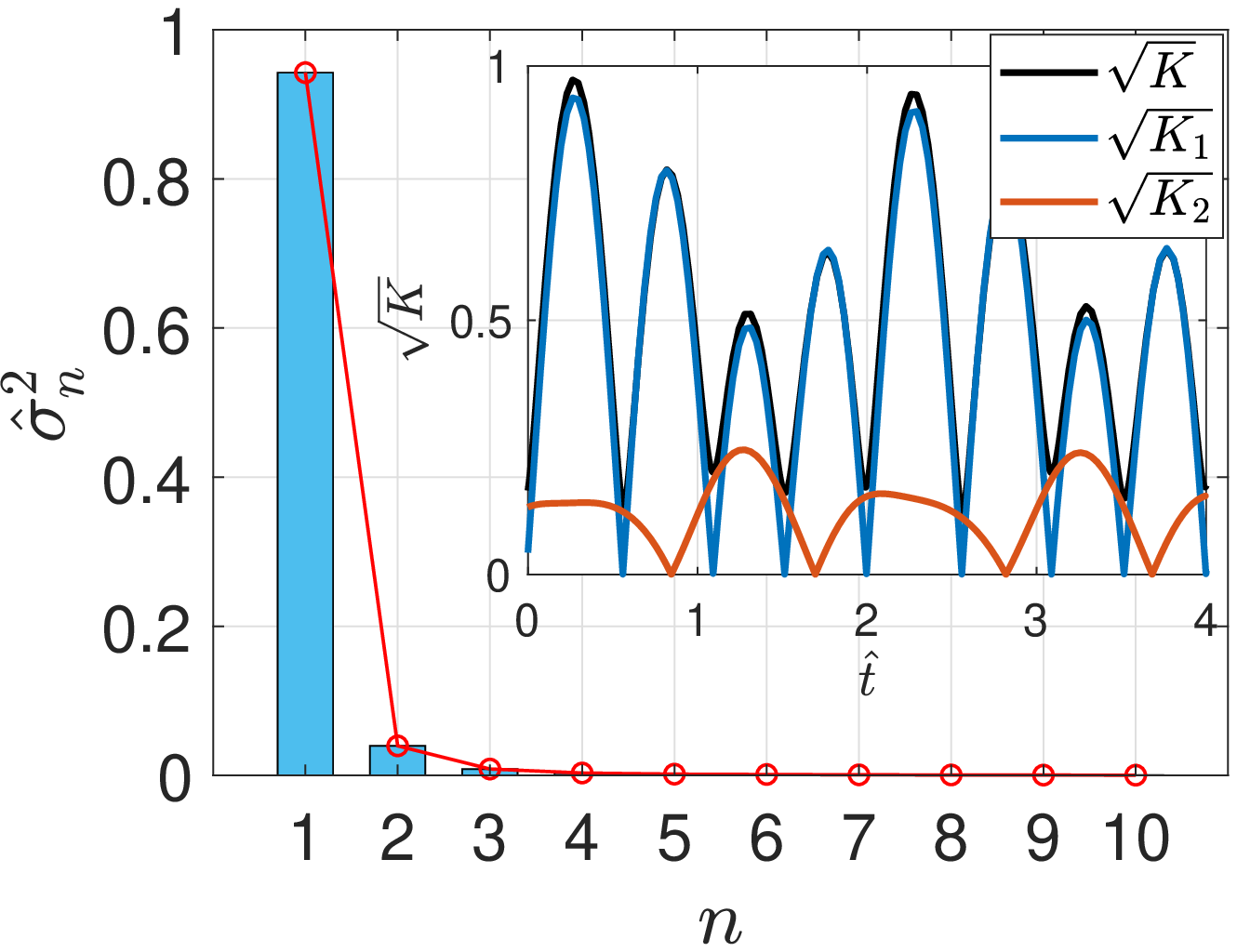}}
    \subfloat[]{\includegraphics[width=54mm,trim=-5 0 30 0, clip]{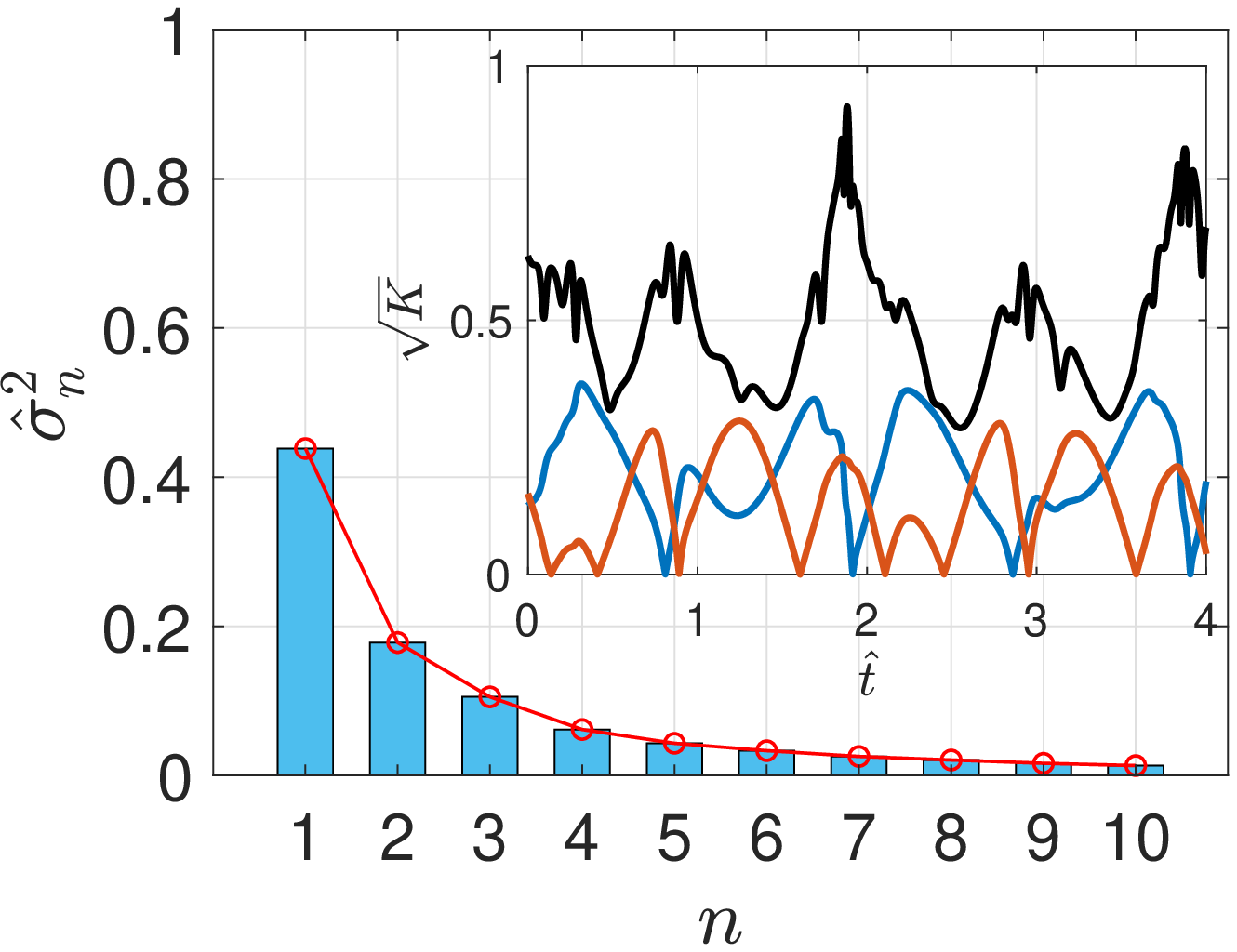}}
    \subfloat[]{\includegraphics[width=54mm,trim=-5 0 30 0, clip]{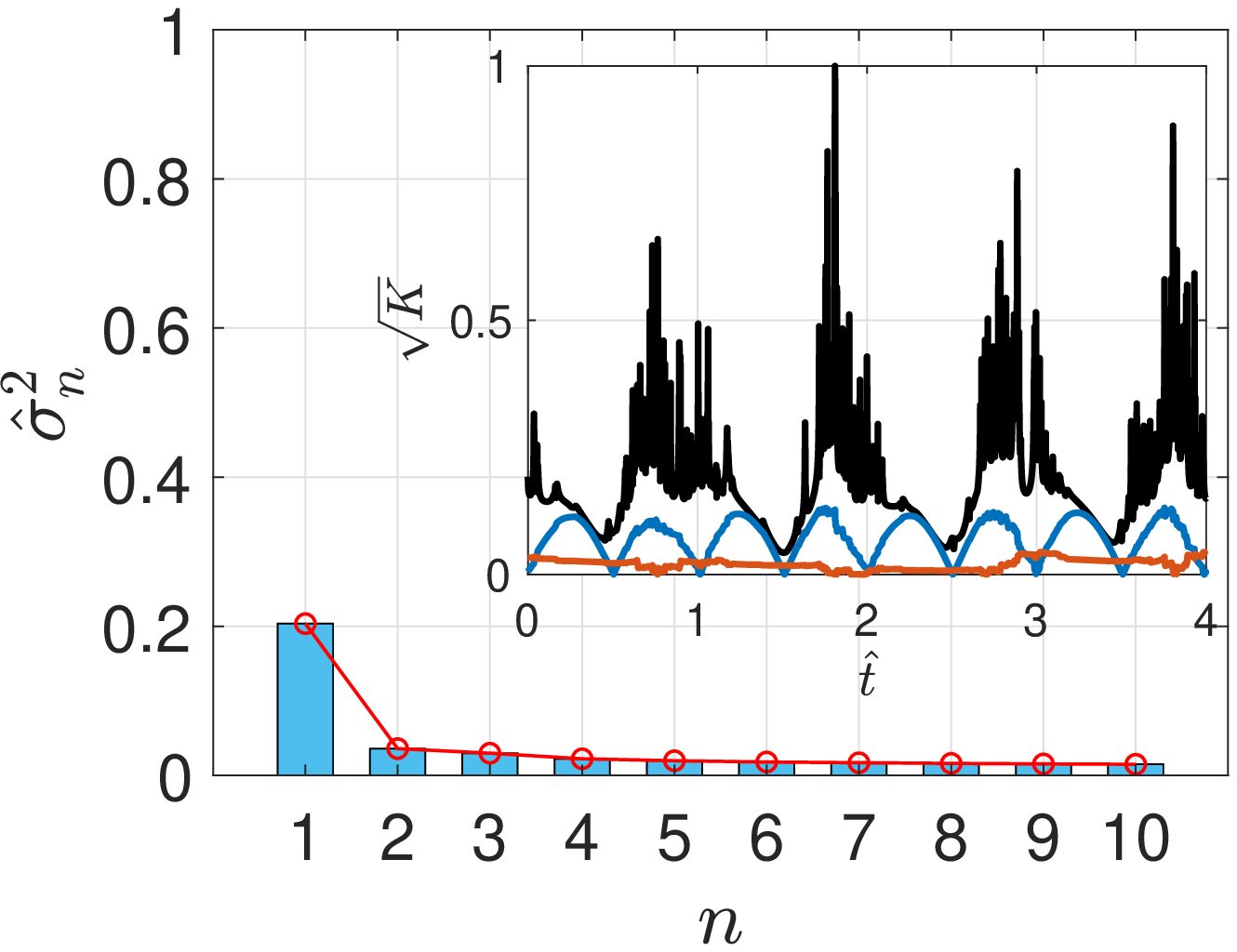}}
    \caption{(a)The first PC-variance ($\hat{\sigma}_1^2$), the normalized von Neumann entropy ($\hat{S}_{vN}$), and the coherence measure ($C$), against the excitation amplitude. (b-d) The PC spectral profiles at $A=2\times10^{-2}$, 1.2, and $20$. Insets show the evolution of the square-root of the normalized total energy ($\sqrt{K}$) and the first and the second PC-variances ($\sqrt{K_1}$ and $\sqrt{K_2}$), during the four periods of excitation, $\hat{t}$ being a non-dimensional time $\hat{t}=tf$. The $y$-axis of each inset is normalized by the maximum value of $\sqrt{K}$.
    }
    \label{fig:s}
\end{figure*}
The coherent dynamics critically depend on the excitation amplitude.
This dependence is best highlighted in the sparse, weakly polydisperse cluster ($B_0=0.5$, $\sigma_d=0.1$).
Fig. \ref{fig:s}a shows the correlations of $\sigma^2_1$, $\hat{S}_{vN}$, and $C$ with $A$.
The relative importance of the first PC, $\hat{\sigma}_1^2$, decays nearly monotonically from 0.9 to 0.2 through three distinct regimes.
For $A\lessapprox0.2$, $\sigma^2_1\approx 1$ meaning the entire energy is captured by the first PC. 
Then $\sigma^2_1$ rapidly decays to 0.4 and stays nearly constant up to $A\approx3$.
At $A>3$, $\sigma^2_1$ decays again and stay nearly constant around 0.2. The decay indicates the decrease in the coherence with increasing $A$ and can be explained by the excitation of the nonlinear oscillations and cavitation triggered at $A\approx1$ and above.
$\hat{S}_{vN}$ has a profile vertically mirrored to $\sigma_1^2$; $\hat{S}_{vN}$ increases from around 0.1 to 0.5 through the three regimes, indicating more partitioning of the energy into multiple PCs and increase of incoherence, by increasing $A$.
The mirrored profiles of $\hat{\sigma}_1^2$ and $\hat{S}_{vN}$ suggest that these parameters are complementary.\\

Remarkably $C$ draws a square-well like profile, where $C\approx1$ for both low ($A<0.2$) and large ($A>5$) amplitudes reaching a minimum value ($C\approx 0.5$)
at $A\approx1$.
This counter-intuitive result suggests that the coherence scaled by $B_d$ is dominant, while the scaling is lost only for the intermediate range at $1<A<5$.
To gain deeper insights, in Fig. \ref{fig:s}b-d we show the PC-variances obtained at $A=2\times10^{-2}$, $1.2$, and $20$, for the first 10 PCs. The insets show the evolution of the square-root of the normalized total energy and those of the relative amplitude of the first and the second PC-variances ($\sqrt{K}$, $\sqrt{K}_1$, and $\sqrt{K}_2$), during the four periods of excitation in statistically stationary states.
As expected, in the linear regime (Fig. \ref{fig:s}b), the first PC occupies nearly the entire energy. $\sqrt{K}_1$ evolves at the fundamental frequency.
In the transition regime (Fig. \ref{fig:s}c), the first PC occupies 40\% of the energy and the rest is partitioned into the sub-dominant PCs with a smooth decay.
Both $\sqrt{K}_1$ and $\sqrt{K}_2$ evolve with similar a-periodic profiles. We interpret that both PCs represent the coherent component of the energy.
Interestingly, in the nonlinear regime (Fig. \ref{fig:s}d) the energy partition is non-smooth; the first PC occupies 20\% of the total energy, while the rest is broadly distributed into the other PCs with a much smaller amplitude. The evolution of $\sqrt{K}$ is noisy, which can be explained by the chaotic response of individual bubbles. The evolution of $\sqrt{K_1}$ is, in contrast, much smoother and periodic, and somewhat resembles that of Fig. \ref{fig:s}b. The evolution of $\sqrt{K_2}$ is a-periodic. The results suggest that, in this regime, only the first PC captures the coherent feature and the others represent the incoherent dynamics as broadband noise.
The resemblance of the evolution of the first PCs in the linear and the nonlinear regimes can explain the recovery of $C$ at $A>5$ in Fig. \ref{fig:s}a. Although the overall dynamics are much noisier in the nonlinear regime, the contribution of the coherent interactions to the system's energy effectively appears only in the first PC in both regimes and therefore the degree of this contribution is scaled by $B_d$ in a similar manner, also implying that the underlying coherence in the nonlinear regime represents the perfect correlation.

\begin{figure}[t]
    \centering
    \subfloat{\includegraphics[width=68mm,trim=0 0 0 0, clip]{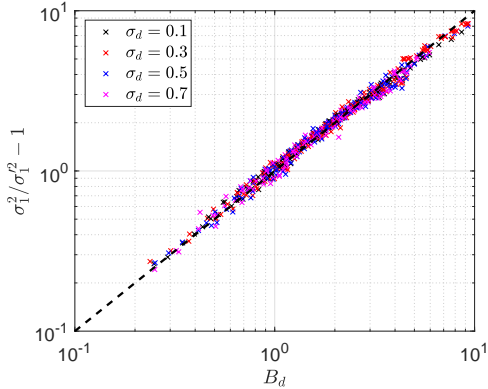}}
    \caption{$\sigma_1^2/{\sigma'}_1^2-1$ against $B_d$, for $O(10^3)$ clusters with various values of $\sigma_d$, $N$, $R_C$, and other physical parameters \citep{Cloud} in the nonlinear regime, excited at $A=20$. The data points are collapsed on the line of $C=1$, indicating that near-perfect correlation is isolated in the first PC and globally scaled by $B_d$, regardless of the parameters.}
    \label{fig:c}
\end{figure}
To assess the universality of the scaling at large $A$, in Fig. \ref{fig:c} we plot $\sigma_1^2/{\sigma'}_1^2-1$ against $B_d$ for $O(10^3)$ clusters with various values of $\sigma_d$, $N$, $R_C$, and other parameters, and constant $A=20$ (2 MPa).
Supplementary material summarizes the parameters \citep{Cloud}.
Surprisingly, the data points are collapsed on the line of $C=1$, meaning that the coherence of these clusters are scaled by $B_d$, regardless of the parameters.
In the linear regime, the scaling is found to hold only for monodisperse clusters \citep{Cloud}, agreeing with the previous theories \citep{dAgostino89,Zeravcic11}.
The results suggest that the scaling is universal for cavitating clusters, which are typically excited at $O(1)$ MPa and above.
Physically speaking, linear mean field represented by $B_d$ originally identified for monodisperse clusters is found here to control the underlying coherence in highly disordered clusters.
This finding also explains the previous success of $B_d$ in characterizing and controlling polydisperse cavitating clusters observed in experiments \citep{Maeda19,Maeda21}.

In conclusion, to address the long-standing question on the coherent dynamics of bubble clusters, we propose a method for unsupervised learning of the many-body physics of interacting bubbles by combining theory and PCA. The method pre-processes bubble dynamics data and extracts coherent dynamical features whose energies are consistent with the R-P equation. Analyzing simulation data sets of clusters under harmonic excitation, we show that the coherence and incoherence are respectively quantified by the PC-variance and the von Neumann entropy.
The PCA also shows that the coherence is lowered by polydispersity and nonlinearity, while under strong excitation near-perfect correlation is isolated in a single coherent mode  globally scaled by $B_d$, regardless of the disorders.
These results highlight the method's use as well as indicate that the underlying coherent dynamics may be universal in cavitating clusters.
For future studies, the scaling argument may provide for a theoretical ground that cavitation and nucleation can be characterized and controlled without measuring individual nuclei.
\newline

Some of the computation presented here utilized the Extreme Science and Engineering Discovery Environment (XSEDE), which is supported by NSF under grant TG-CTS190009.
K.M. conceived and designed the study, developed methodology, and performed data processing and analysis. Both authors contributed to data generation and the manuscript.

\bibliographystyle{science}

\providecommand{\noopsort}[1]{}\providecommand{\singleletter}[1]{#1}%

\newpage
\begin{center}
\section*{\Large{Supplementary information}} 
\end{center}
\setcounter{section}{0}
\setcounter{figure}{0}
\renewcommand{\figurename}{Figure S}
\section{Keller-Miksis equation for multiple bubbles}
In this section, we provide details of the formulation of the Keller-Miksis (K-M) equation for multiple bubbles that is used to generate the Lagrangian bubble dynamics data.

Various extensions of the Rayleigh-Plesset equation to a system of ODEs which describe the dynamics of multiple bubbles are available [S1-4].
As a variant of such extensions, we employ the K-M equation for multiple bubbles to account for liquid compressiblity which becomes important for the dynamics of cavitating bubbles.
To begin with, we recall the general formulation for interacting spherical bubbles in weakly compressible liquid with arbitrary inter-bubble distances, as provided in Appendix 2.4 of [S5].
With $c$ being the sound speed in liquid, the radial evolution of bubble $i$ is described as

\begin{equation}
    \ddot{R}_i\left(R_i\left(1-\frac{\dot{R}_i}{c}\right)\right)+\frac{3}{2}R^2_i\left(1-\frac{\dot{R}_i}{3c}\right)=F^*+I^*,
    \label{eqn:KM}
\end{equation}
where $F^*$ and $I^*$ represent the forcing due to the external potential and the inter-bubble interaction, respectively, and are expressed as
\begin{equation}
    F^*=\frac{\partial\phi_\infty}{\partial t}\left(1-\frac{\dot{R}_i}{c}\right)+\frac{R_i}{c}\frac{\partial^2\phi_\infty}{\partial t^2}+H_i\left(1+\frac{\dot{R}_i}{c}\right)+\frac{R_i\dot{H}_i}{c},
\end{equation}
and
\begin{align}
    I^*
    =&\sum^N_{j\ne i}\left[\left(1+\frac{\dot{R}_i}{c}\right)\frac{\partial\phi_j(R_i)}{\partial t}\right]
    +\frac{R_i}{c}\sum^N_{j\ne i}
    \left[\left(1+\frac{\dot{R}_j(t')}{c}\right)\left(1-\frac{\dot{R}_j}{c}\right)\frac{\partial^2\phi_j(t'-R_j(t')/c)}{\partial t'^2}\right]\\
    &-\sum^N_{j\ne i}\left[\left(1+\frac{\dot{R}_j(t')}{c}\right)\frac{\partial\phi_j(t'-R_j(t')/c)}{\partial t'}\frac{R_i}{R_j}\frac{\dot{R}_j}{c}\right].
\end{align}
$\phi_\infty$ is the velocity potential of liquid at infinity, and $\phi_i(R_i)$ and $H_i$ are the potential and the enthalpy of liquid evaluated at the surface of bubble $i$.
$t'$ is the retarded time defined as $t'=t-(d_{ij}-R_j)/c$, where $(d_{ij}-R_j)/c$ represents the travel time for the pressure wave to reach bubble $i$ from the surface of the bubble $j$ and $d_{ij}$ is the distance between the centers of those bubbles.
In the sparse limit, the equation recovers the original Keller-Miksis equation [S6].

To close the equations, several relations are considered.
Using the Bernoulli's equation, the potential for the bubble $i$ can be expressed as
\begin{align}
    \frac{\partial\phi_i(R_i)}{\partial t}=-\left(\frac{1}{2}\dot{R}^2_i+H_i+\sum^N_{j\ne i}\frac{\partial\phi_j(R_i)}{\partial t}+\frac{\partial\phi_\infty}{\partial t}\right).
\end{align}
The velocity potentials at the bubble $i$ and bubble $j$ satisfy the following relation.
\begin{align}
    \frac{\partial\phi_i(R_i)}{\partial t}=\frac{R_j(t')}{d_{ij}}\frac{\partial\phi_j(t'-R_j/c)}{\partial t'}.
\end{align}

The enthalpy and the potential derivative are approximated as
\begin{equation}
    H_i\approx\frac{p_i-p_0}{\rho_{l,0}}.
\end{equation}
\begin{equation}
    \frac{\partial\phi_\infty}{\partial t}\approx\frac{p_\infty-p_0}{\rho_{l,0}}.
\end{equation}

Finally, $p_\infty$ is the background pressure known a-priori (the pressure at the location of the bubble excluding the contributions of the bubbles oscillations, typically equivalent with the excitation pressure imposed by external sources).

This system of ODEs, however, involves time delay (delayed differential equations: DDE) and simulations can become cumbersome.
Therefore, we newly simplify the system by invoking further approximations.
First, consider that the bubble cluster size is smaller than the characteristic length-scale of the pressure wave in the field, $\lambda_c$. The inter-bubble distance in the cluster is naturally smaller than $\lambda_c$:
\begin{equation}
    d_{ij}-R_j\ll\lambda_c.
\end{equation}
Dividing both sides with $c$,
\begin{equation}
    \frac{d_{ij}-R_j}{c}=t-t'\ll\frac{\lambda_c}{c}=\frac{1}{f_c}=T_c,
\end{equation}
where $f_c$ and $T_c$ are the characteristic frequency and the period of the wave in the field.
Therefore, the difference between $t$ and $t'$ is much smaller than the characteristic timescale of the dynamics.
Given this knowledge, we approximate that $t'\approx t$.
Second, we neglect the terms in the order of $(\dot{R}/c)^2$.
Using these approximations, we can simplify $F^*$ and $I^*$ as
\begin{equation}
    F^*=\frac{p_i-p_\infty}{\rho_l}\left(1+\frac{\dot{R}_i}{c}\right)+\frac{R_i}{\rho_lc}\frac{\partial(p_i-p_\infty)}{\partial t}
    \label{eqn:F}
\end{equation}
and
\begin{equation}
    I^*=-\sum^N_{j\ne i}\frac{R_j}{d_{ij}}\left(\frac{1}{2}\dot{R}^2_j+\ddot{R}_j\left(R_j\left(1-\frac{\dot{R}_j}{c}\right)\right)+\frac{3}{2}\dot{R}^2_j\left(1-\frac{\dot{R}_j}{3c}\right)-\frac{p_j-p_\infty}{\rho_l}\frac{\dot{R}_j}{c}-\frac{R_j}{\rho_l c}\frac{\partial(p_j-p_\infty)}{\partial t}\right).
    \label{eqn:I}
\end{equation}

{For harmonic excitation,
\begin{equation}
    p_\infty=p_a\mathrm{sin}(\omega t). \label{eqn:pinf}
\end{equation}}
We use polytropic law to describe the pressure of the gas inside each bubble [S7].
{
\begin{equation}
    p_i=p_{i,0}\left(\frac{R_{i,0}}{R_i}\right)^{3\gamma}, \label{eqn:pi}
\end{equation}
where $\gamma$ is the constant polytropic exponent.
Eqns (\ref{eqn:KM}), (\ref{eqn:F}) and (\ref{eqn:I}), together with relations (\ref{eqn:pinf}) and (\ref{eqn:pi}) provide a complete system of ODEs for the radius of $N$ interacting bubbles, which can be readily solved with a given initial condition using a standard time integration method.}

\section{Summary of the parameters}

\begin{table}[h]
\centering
\begin{tabular}{ccccc}
\hline\hline
\textrm{$N$}&
\textrm{$R_c$ [mm]}&
\textrm{$R_{ref}$ [$\mu$m]}&
\textrm{$\sigma_d$}&
\textrm{$f$ [kHz]}\\
\hline
16 & 0.33-2.0 & 10 & 0.1-0.7 & 500\\
32 & 0.33-2.0 & 10 & 0.1-0.7 & 500\\
64 & 0.33-2.0 & 10 & 0.1-0.7 & 500\\
128 & 0.33-2.0 & 10 & 0.1-0.7 & 500\\
256 & 0.33-2.0 & 10 & 0.1-0.7 & 500\\
512 & 0.8-2.0 & 10 & 0.1-0.7 & 500\\
64 & 0.33-2.0 & 5 & 0.1-0.7 & 500\\
64 & 0.33-2.0 & 10 & 0.1-0.7 & 250\\
\hline\hline
\end{tabular}
\caption{\label{tab:table1}%
Summary of the parameters used for the simulation of clusters plotted in figure 4 of the main manuscript.
}
\label{t1}
\end{table}
Table \ref{t1} summarizes a set of parameters for the clusters plotted in figure 4 of the main manuscript.
The number of bubbles $N$, cluster radius $R_c$, reference bubble radius $R_{ref}$, polydispersity measure $\sigma_d$, and the forcing frequency were varied.

\section{Scaling in the linear and the transition regimes}
In order to assess the scaling of the coherence in the linear and the transition regimes of bubble clusters, we simulate various parameters of clusters with a weak ($A=0.01$) and an intermediate amplitude ($A=2$) of pressure waves. The parameters used for these simulations are summarized in table \ref{t2}.
\begin{table}[h]
\centering
\begin{tabular}{ccccc}
\hline\hline
\textrm{$N$}&
\textrm{$R_c$ [mm]}&
\textrm{$R_{ref}$ [$\mu$m]}&
\textrm{$\sigma_d$}&
\textrm{$f$ [kHz]}\\
\hline
16 & 0.33-2.0 & 10 & 0.1-0.7 & 500\\
32 & 0.33-2.0 & 10 & 0.1-0.7 & 500\\
64 & 0.33-2.0 & 10 & 0.1-0.7 & 500\\
128 & 0.33-2.0 & 10 & 0.1-0.7 & 500\\
\hline\hline
\end{tabular}
\caption{\label{tab:table2}%
Summary of the parameters used for the simulation of clusters plotted in Supplementary figure 1.
}
\label{t2}
\end{table}

\begin{figure}[t]
    \centering
    \subfloat{\includegraphics[width=64mm,trim=0 0 0 0, clip]{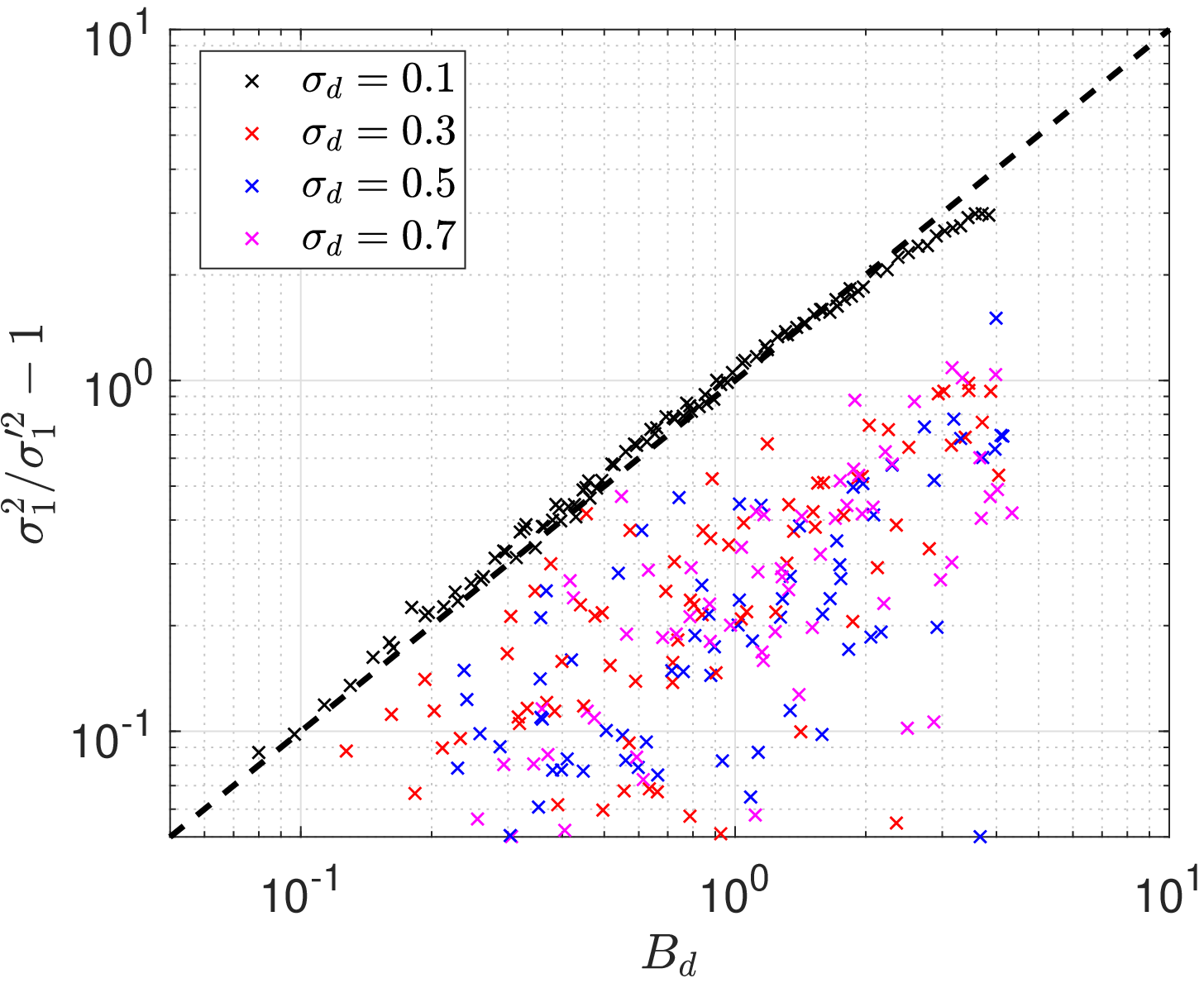}}
    \subfloat{\includegraphics[width=64mm,trim=0 0 0 0, clip]{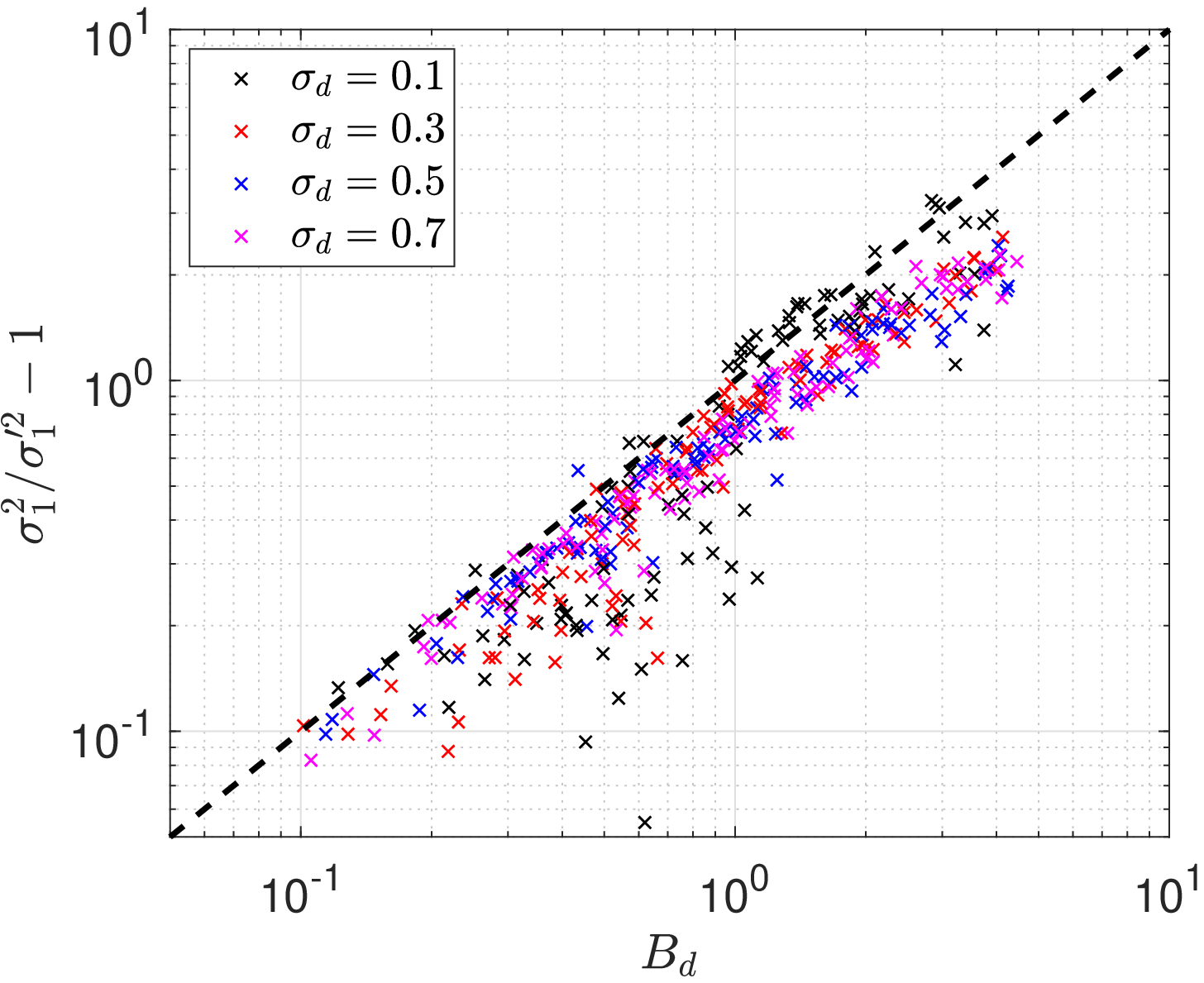}}
    \caption{$\sigma_1^2/{\sigma'}_1^2-1$ as a function of $B_d$, for clusters with various parameters. (a) $A=0.01$ and (b) 2.0.}
    \label{fig:c2}
\end{figure}
Figure \ref{fig:c2} shows $\sigma_1^2/{\sigma'}_1^2-1$ as a function of $B_d$ for these clusters.
For the case of the weak amplitude, bubble oscillations are in a linear regime and $B_d\approx B_0$.
The data points are collapsed on the coherence line ($C=1$) for $\sigma_d=0.1$, while data are scattered for the other values of $\sigma_d$. This result is expected as $B_d$ was originally defined to scale the coherence of the monodisperse, perfectly correlated bubbles in the linear regime.
For the intermediate amplitude case, the bubble oscillations are nonlinear. The data points are scattered from the coherence line, regardless of the value of $\sigma_d$. In this regime, the result indicates that the coherence is lost due to the nonlinear dynamics, regardless of polydispersity.

\section*{References}
\footnotesize{
\noindent[S1] A.A. Doinikov. Mathematical model for collective bubble dynamics in strong ultrasound fields. {\it{J. Acoust. Soc. Am.}}, 116(2):821–827, 2004.\\
\noindent[S2] H. Takahira, T. Akamatsu, and S. Fujikawa. Dynamics of a cluster of bubbles in a liquid: Theoretical analysis. {\it{JSME Int. J. Ser. B Fluids Therm. Eng.}}, 37(2):297–305, 1994.\\
\noindent[S3] Y.A. Ilinskii, M.F. Hamilton, and E.A. Zabolotskaya. Bubble interaction dynamics in Lagrangian and Hamiltonian mechanics. {\it{J. Acoust. Soc. Am.}}, 121(2):786–795, 2007.\\
\noindent[S4] K. Yasui, Y. Iida, T. Tuziuti, T. Kozuka, and A. Towata. Strongly interacting bubbles under an ultrasonic horn. {\it{Phys. Rev. E}}, 77(1):016609, 2008.\\
\noindent[S5] D. Fuster and T. Colonius. Modelling bubble clusters in compressible liquids. {\it{J. Fluid Mech.}}, 688:352–389, 2011.\\
\noindent[S6] J.B. Keller and M. Miksis. Bubble oscillations of large amplitude. {\it{J. Acoust. Soc. Am.}}, 68(2):628–633, 1980.\\
\noindent[S7] C.E. Brennen. Cavitation and bubble dynamics. Cambridge University Press, 2014.
}

\end{document}